\documentclass[12pt,letterpaper]{JHEP3}
\pdfoutput=1
\usepackage{cite}
\usepackage{epsfig}

\title{Complementarity in the Multiverse}

\author{Raphael Bousso\\
  Center for Theoretical Physics, Department of Physics\\
  University of California, Berkeley, CA 94720-7300, U.S.A.\\
  {\em and}\\
  Lawrence Berkeley National Laboratory, Berkeley, CA 94720-8162,
  U.S.A.}

\abstract{In the multiverse, as in AdS, light-cones relate bulk points
  to boundary scales.  This holographic UV-IR connection defines a
  preferred global time cut-off that regulates the divergences of
  eternal inflation.  An entirely different cut-off, the causal patch,
  arises in the holographic description of black holes.  Remarkably, I
  find evidence that these two regulators define the same probability
  measure in the multiverse. Initial conditions for the causal patch
  are controlled by the late-time attractor regime of the global
  description.}

\begin{document}

\section{Introduction}
\label{sec-intro}

\begin{figure*}
\begin{center}
\includegraphics[scale = .5]{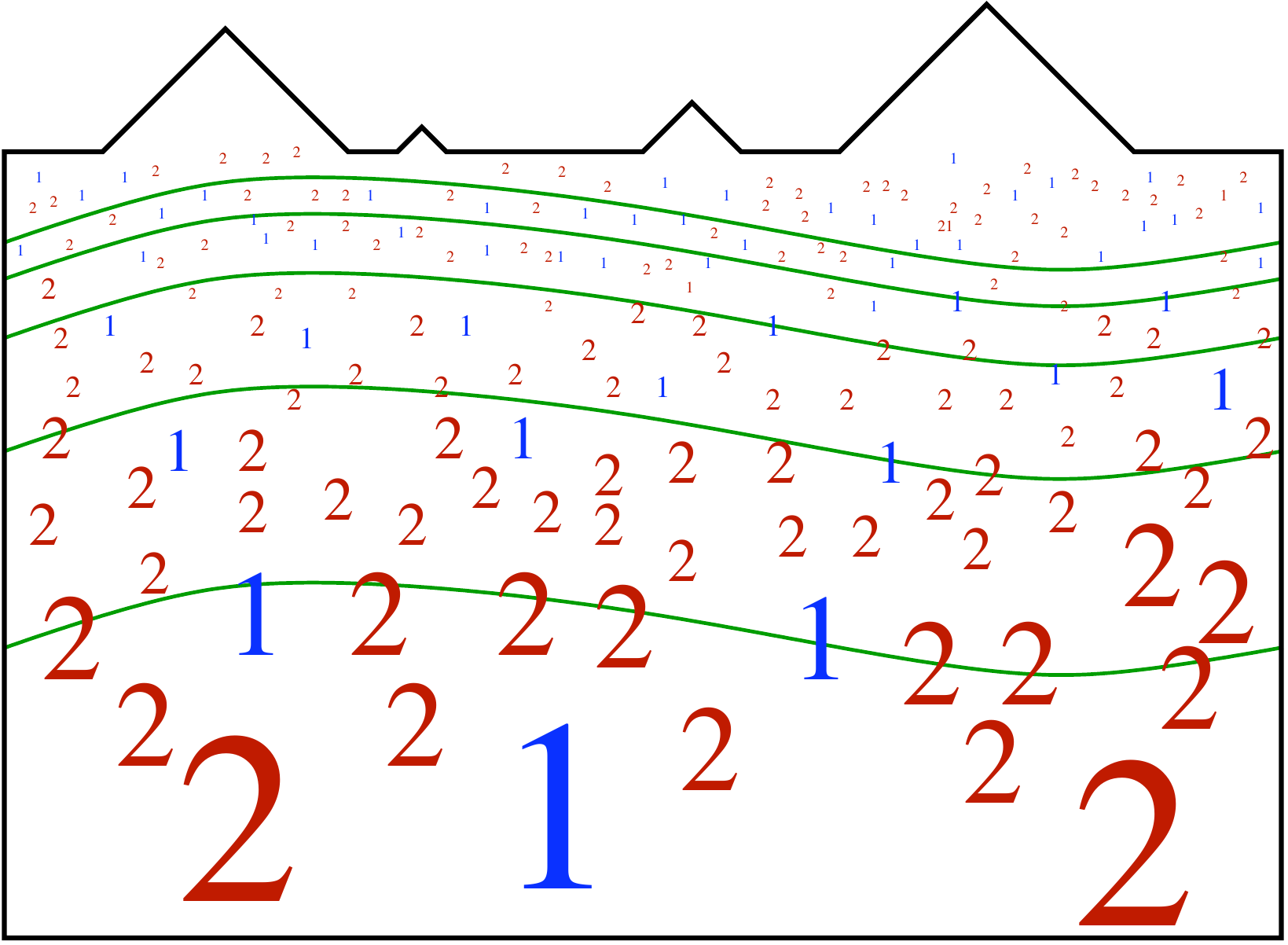}
\end{center}
\caption{A global cut-off on the multiverse.  The curvy lines represent
  hypersurfaces of constant $t$, with $t\to\infty$ at the top of the
  diagram. The relative probability of events 1 and 2 is defined by
  $p_1/p_2=\lim_{t\to\infty} N_1(t)/N_2(t)$, where $N_i(t)$ is the
  number of times a given event has occurred by the time $t$.  This
  definition is ambiguous, because it depends on the choice of
  foliation, i.e., of the time coordinate $t$.}
\label{fig-globalcut}
\end{figure*}

In global approaches to the measure problem of eternal inflation, one
constructs a time variable $t$ that foliates the multiverse
(Fig.~\ref{fig-globalcut}).  The relative probability for observations
$i$ and $j$ is defined by
\begin{equation}
  \frac{p_i}{p_j}=\lim_{t\to\infty}\frac{N_i(t)}{N_j(t)}~,
  \label{eq-nt}
\end{equation}
where $N_i(t)$ is the number of times that $i$ is observed somewhere
in the multiverse prior to the time t.  In the late time limit, both
$N_i$ and $N_j$ diverge---this, of course, is the origin of the
measure problem.  But one expects their ratio to remain finite and to
converge to a well-defined relative probability.

This prescription is ambiguous.  There are many ways to define a time
variable, and the probabilities in Eq.~(\ref{eq-nt}) are highly
sensitive to its choice.  In a beautiful recent paper, Garriga and
Vilenkin~\cite{GarVil08} have outlined a novel approach to identifying
a preferred foliation.  In the spirit of the AdS/CFT
correspondence~\cite{Mal97}, the idea is to specify a short-distance
cut-off $\epsilon$ on the future boundary of the multiverse.  A
conjectured duality between a boundary theory and the bulk physics
should relate this ultraviolet cut-off to a late-time cut-off $t$ in the
bulk, with the property that $t\to\infty$ as $\epsilon\to 0$.

\begin{figure*}
\begin{center}
\includegraphics[scale = .7]{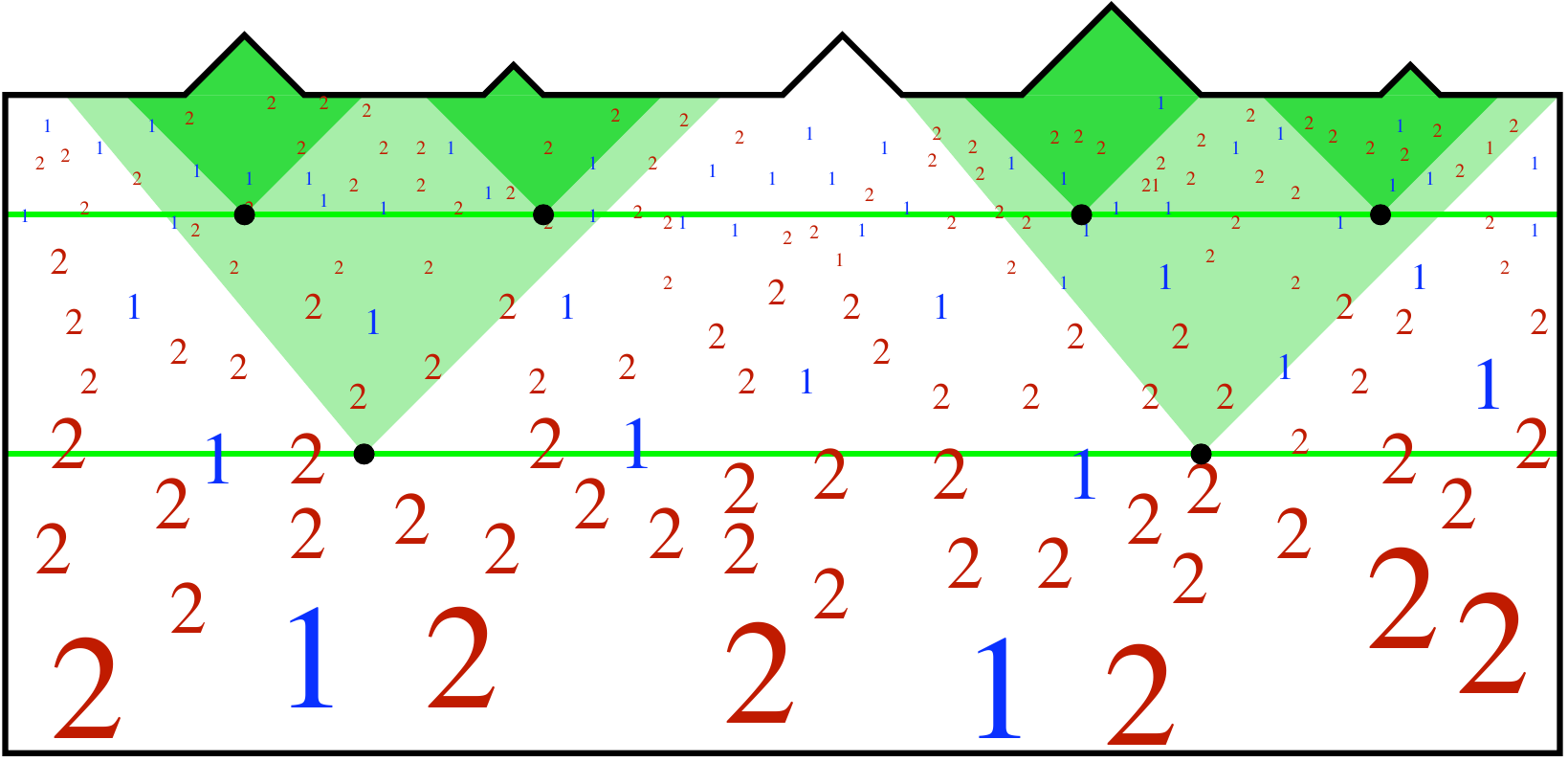}
\end{center}
\caption{The future light-cone of any event (black dots) defines a
  scale on the asymptotic boundary of the multiverse (top edge).
  (Like in AdS, this scale is physically infinite but can be regulated
  as described in the text.)  Conversely, constant light-cone size on
  the boundary defines a hypersurface of constant ``light-cone time''
  in the bulk (green horizontal lines).  Taking the boundary scale to
  zero generates a preferred foliation of the multiverse.  Remarkably,
  the resulting global probability measure is equivalent to the causal
  patch measure (below).}
\label{fig-lightconecut}
\end{figure*}

In order to realize this idea, one needs to construct a precise
relation between the boundary scale $\epsilon$ and some finite portion
of the bulk.  One might be concerned that this task is no less
ambiguous than the original problem of picking a time slicing.  In
this paper I will argue, however, that the UV-IR relation of the
multiverse, like that of AdS/CFT, is unambiguously determined by
causality (Fig.~\ref{fig-lightconecut}).  I will thus construct a
preferred foliation $t$ and, via Eq.~(\ref{eq-nt}), the associated
probability measure.

\begin{figure*}
\begin{center}
\includegraphics[scale = .7]{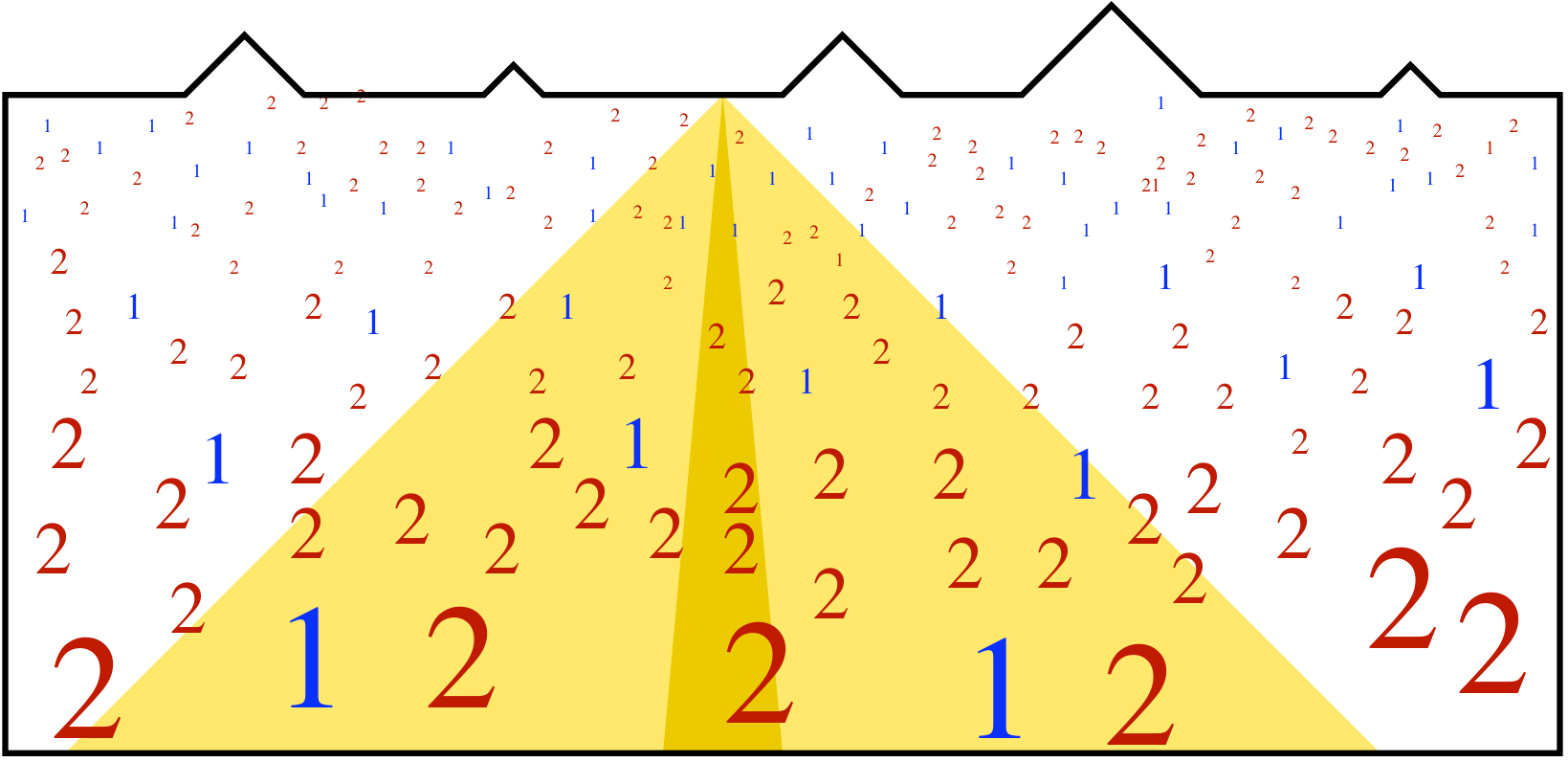}
\end{center}
\caption{The causal patch measure abandons the global description of
  the multiverse. Relative probabilities are defined by ratios of the
  expected numbers of events of different types taking place within
  the causal past of a single worldline (shaded region), averaging
  over possible histories and initial conditions.  (The darker-shaded
  narrow triangle is discussed in Sec.~5.)}
\label{fig-causalcut}
\end{figure*}

A different response to the ambiguities of Eq.~(\ref{eq-nt}) has been
to abandon the global description of the multiverse.  Motivated by
black hole complementarity~\cite{SusTho93},
Refs.~\cite{Bou06,BouFre06a} advocated that no more than one causally
connected region (causal diamond, or causal patch) should be
considered.  This led to the causal patch measure~\cite{Bou06}
(Fig.~\ref{fig-causalcut}), which has had considerable
phenomenological success~\cite{BouYan07,BouHar07}.

The global and causal patch approaches appear to be radically
different.  Remarkably, I will find that they make the same
predictions: The global probability measure (\ref{eq-nt}) arising from
the ultraviolet boundary cut-off $\epsilon$ is equivalent to the causal
patch measure.  Thus, the global and causal patch viewpoints can be
reconciled.  They are not contradictory but complementary, or dual.

\paragraph{Outline} Before considering the multiverse, I begin in
Sec.~\ref{sec-ads} by reviewing the UV-IR connection of AdS/CFT.  I
show that the geometric relation between bulk points and boundary
scales~\cite{SusWit98} is determined by causality and can be
constructed without detailed knowledge of the duality.  This
construction has a natural analogue in the multiverse, which I present
in Sec.~\ref{sec-multi}.  The ``light-cone time'' $t$ associated to a
bulk point $p$ is given by
\begin{equation}
t(p)=-\frac{1}{3}\log\epsilon(p)~,
\end{equation}
where $\epsilon(p)$ is the (suitably defined) volume of the
chronological future of $p$ on the boundary.

In Ref.~\cite{GarVil08}, Garriga and Vilenkin propose a different
UV-IR connection in the multiverse.\footnote{The general idea of a
  boundary cut-off~\cite{GarVil08} forms the basis of the present paper
  and is left untouched.  I depart from Garriga and
  Vilenkin~\cite{GarVil08} only in that I argue that a careful analogy
  with AdS/CFT leads to a different implementation of the novel
  approach they have proposed.} This connection involves an extra
parameter $\lambda$ that has no analogue in AdS/CFT, and which is
somewhat loosely defined.  Here, I will treat $\lambda$ as a free
parameter that can be used to clarify relations between different
measures.  If $\lambda$ is constant, then the Garriga-Vilenkin
connection reduces to the well-known scale factor
measure~\cite{DGSV08} (see also
Refs.~\cite{LinLin94,GarLin94,GarLin94a,GarLin95,Lin06}).  In
Sec.~\ref{sec-gv}, I generalize this result to the case where
$\lambda$ depends on the bulk point $p$.  Under rather weak
assumptions, the resulting class of measures is related to the scale
factor measure in a simple and specific way.

This result will be useful in Sec.~\ref{sec-equiv}, where I
investigate the probability measure defined by the light-cone time
$t$.  I show that the light-cone construction of Sec.~\ref{sec-multi}
can be regarded as selecting a particular choice of $\lambda(p)$, in
the language of Sec.~\ref{sec-gv}.  This makes it possible to quantify
precisely how the light-cone measure differs from the scale factor
measure.  I show that in a wide class of multiverse regions, this
difference is precisely the difference between the causal patch
measure and the scale factor measure.  Thus, the light-cone cut-off is
equivalent to the causal patch measure.  This result is discussed in
Sec.~\ref{sec-discussion}.

\section{The UV-IR relation of AdS/CFT from light-cones}
\label{sec-ads}

In this section, I will show that the geometric aspects of the
bulk-boundary relation of AdS/CFT are determined by causality.  The
AdS/CFT correspondence is the statement that quantum gravity in
asymptotically Anti-de~Sitter spacetimes is nonperturbatively defined
by a conformal field theory that can be thought of as living on the
spatial boundary of Anti-de~Sitter space.  If the AdS curvature radius
is large, the spacetime will be accurately described by general
relativity as well.  In this regime, the AdS/CFT correspondence
implies a highly nontrivial equivalence between the boundary quantum
field theory and the classical dynamics in the bulk.

\begin{figure*}
\begin{center}
\includegraphics[scale = 1]{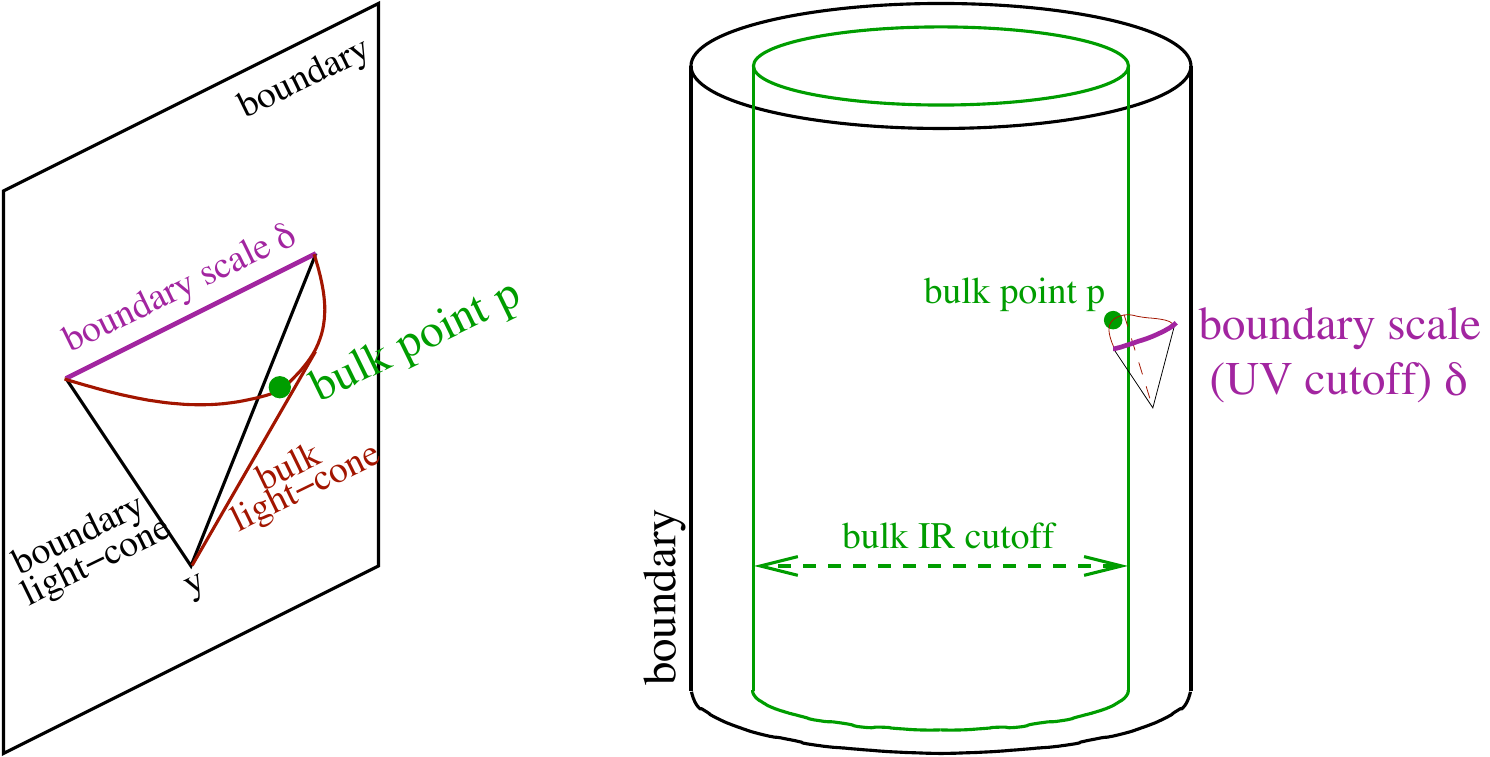}
\end{center}
\caption{Every point $p$ in Anti-de Sitter space is associated to a
  boundary scale $\delta$.  This relation is completely determined by
  causality (left; only a small portion of the boundary is shown).  It
  implies the well-known UV-IR connection of the AdS/CFT
  correspondence (right; here the global geometry is shown).}
\label{fig-adsbb}
\end{figure*}

The metric of AdS in $D$ space time dimensions is
\begin{equation}
  ds^2=\frac{L_{\rm AdS}^2}{\cos^2\rho}
  \left(-d\tau^2 + d\rho^2 + \sin^2\rho~d\Omega_{D-2}^2\right)~,~~~
  0\leq\rho< \frac{\pi}{2}~,
\label{eq-ads}
\end{equation}
(I will not include the extra product space factors that appear in
supergravity solutions, such as AdS$_5\times \mathbf{S}^5$, because
they play no role in this analysis.)  The proper area-radius of
spheres in the bulk is related to the coordinate $\rho$ by
\begin{equation}
r=L_{\rm AdS} \tan\rho~.
\label{eq-rrho}
\end{equation}
By dropping the conformal factor $L_{\rm AdS}^2/\cos^{2}\rho$, one
sees that the conformal boundary of AdS has the structure
$\mathbf{R}\times \mathbf{S}^{D-2}$.  It can be thought of as a unit
$D-2$ sphere sitting at $\rho=\pi/2$ at all times $\tau$
(Fig.~\ref{fig-adsbb}, right).

Both AdS and the CFT contain an infinite number of degrees of freedom.
Now, let us introduce an infrared cut-off in the bulk, keeping a
spatially finite portion of AdS space, $r<r_{\rm c}$, and removing points
closer to the boundary.  I will assume that $r_{\rm c}\gg L_{\rm AdS}$.  The
surviving region has finite information content, limited by the
entropy of the largest black hole that can fit: $S\sim r^{D-2}$ in $D$
spacetime dimensions.

A finite portion of the Hilbert space of the conformal field theory
should suffice to describe this region.  Indeed, there is overwhelming
evidence that the truncation of the bulk corresponds to an ultraviolet
cut-off in the CFT, at a distance scale
\begin{equation}
  \delta_{\rm c} \sim \frac{L_{\rm AdS}}{r_{\rm c}}
\label{eq-dr}
\end{equation}
on the unit sphere.  (Note that $\delta_{\rm c}\ll 1$.)  For example,
correlators begin to differ from the conformal scaling at this
distance.  Moreover, the IR bulk cut-off regulates the divergent energy
of a string stretched across the bulk in precisely as the UV boundary
cut-off regulates its boundary dual (a point charge).  Crucially, the
maximum information content of the UV-regulated boundary theory is of
order $L_{\rm AdS}^{D-2}/\delta^{D-2}$, which matches the entropy
$r^{D-2}$ of the largest black hole allowed by the IR bulk
cut-off~\cite{SusWit98}.

In the above examples, the consistency of the UV-IR relation,
Eq.~(\ref{eq-dr}), depends on detailed properties of the AdS/CFT
correspondence in string theory.  In hindsight, however, the relation
itself could have been obtained from a geometric analysis alone, under
two rather weak assumptions: (1) Bulk and boundary theories are both
causal; and (2) As a localized bulk excitation approaches a boundary
point $p$, its dual boundary degrees of freedom also become localized
at $p$.  Both of these assumptions hold true in AdS/CFT, but they are
of course far weaker than the full duality.\footnote{A closely related
  argument was presented in Ref.~\cite{BouRan01} in a different
  context; see in particular Sec.~5 therein.}

Let us define a coordinate $\delta\equiv L_{\rm AdS}/r$.  By
Eq.~(\ref{eq-rrho}), $\delta\approx\frac{\pi}{2}-\rho$ near the
boundary, where we can approximate the AdS metric as
\begin{equation}
  ds^2=\frac{1}{\delta^2}
  \left(-d\tau^2 + d\delta^2+ d {\mathbf x}^2\right)~,~~~
    0\leq\delta\ll 1~,~~~|{\mathbf x}|\ll 1 ~;
\end{equation}
see Fig.~\ref{fig-adsbb} (left).  In otherwise empty AdS space,
consider an excitation localized at $\tau=0$, ${\mathbf x}=0$ on the
boundary, $\delta=0$.  As this excitation propagates into the bulk,
causality requires that it remain localized within the light-cone,
$\delta^2+|{\mathbf x}|^2\leq\tau^2$.  As it propagates on the
boundary, causality of the CFT requires that it remain within
$|{\mathbf x}|^2\leq\tau^2$.  It will suffice to consider values of
$\tau\ll 1$.

Now let us impose an infrared cut-off $r\lesssim L_{\rm
  AdS}/\bar\delta_{\rm c}$ on the bulk, ignoring all points with
$\delta\lesssim\bar\delta_{\rm c}$.  The above excitation first enters
the surviving portion of the bulk at the time
$\tau_1\sim\bar\delta_{\rm c}$, at the point ${\mathbf x} =0$.  Let us
also impose an ultraviolet cut-off $|\Delta x|\gtrsim\delta_{\rm c}$ on
the boundary, ignoring modes smaller than $\delta_{\rm c}$.  Then the
above excitation is first resolved by the boundary theory at the time
$\tau_2\sim\delta_{\rm c}$, when the boundary light-cone becomes
larger than $\delta_{\rm c}$.  Unless $\tau_1=\tau_2$, there will be
times when the excitation is described by the boundary theory but
absent from the bulk, or present in the bulk but not in the boundary
theory.  For the two descriptions (bulk with IR cut-off, boundary with
UV cut-off) to be approximately equivalent, it follows that one must
choose $\delta_{\rm c}\sim\bar\delta_{\rm c}$.  This implies the UV-IR
relation of Eq.~(\ref{eq-dr}).%
\footnote{It may appear that this conclusion depends on the way the
  boundary time coordinate $\tau$ is extended into the bulk.  (Suppose
  we used a coordinate $\hat\tau$ that agrees with $\tau$ on the
  boundary but not away from it.  Then we might obtain a different
  relation between $\delta$ and $\bar\delta$.)  In fact, there is no
  ambiguity.  A short-distance cut-off breaks the symmetries of the
  boundary and picks out a preferred time coordinate $\tau$.  This
  coordinate must be extended to the bulk by assigning to any point
  $p$ the value $\tau=\frac{1}{2}[\tau_+(p)+\tau_-(p)]$, where
  $\tau_+$ ($\tau_-$) is the earliest (latest) value of $\tau$ at
  which the future (past) light-cone of $p$ reaches the boundary.
  (Any other choice would break time-reversal invariance.)
  Alternatively, one could make the construction of $\delta(p)$
  explicitly independent of the choice of bulk time, by introducing a
  second excitation localized on the boundary at the time $\tau_{\rm
    c}>0$ and considering its past-directed causal propagation into
  the bulk.  Then $\delta_{\rm c}=\tau_{\rm c}/2$ is unambiguously the
  innermost bulk point at which both signals are present, and also the
  largest boundary distance on which both signals simultaneously have
  support.}

Let us summarize this result.  Each point (event) $p$ in Anti-de
Sitter space can be associated with a length scale $\delta(p)$ on its
spatial boundary, such that $\delta\to 0$ as $p$ approaches the
boundary.  Given a point $p$, $\delta(p)$ can be constructed by
considering the causal future $I^+(y)$ of a boundary point $y$ chosen
so that $p$ is barely contained in $I^+(y)$ (i.e., $\partial I^+(y)$
passes through $p$).  Then $\delta(p)$ is the spatial size of the
causal future of $y$ on the boundary, at the time when the light-cone
reaches $p$ (Fig.~\ref{fig-adsbb}).  This relation defines a UV-IR
connection: The boundary theory with a UV cut-off $\delta_{\rm c}$ describes
the portion of the bulk whose points satisfy $\delta(p)>\delta_{\rm c}$.  We
thus see that the UV-IR relation of AdS/CFT is determined by causality
alone and requires no detailed knowledge of the rich structures on
both sides of the duality.

\section{The UV-IR relation of the multiverse}
\label{sec-multi}

I have shown that null hypersurfaces uniquely relate the bulk and
boundary of AdS, and that they constrain the UV-IR connection quite
independently of the details of the correspondence.  Though neither
fact, perhaps, is widely appreciated, neither is surprising: In
general spacetimes, holography can {\em only\/} be defined in terms of
null hypersurfaces~\cite{CEB2}, since this is the only setting in
which entropy is always bounded by area~\cite{CEB1,Bou02}.  In the
multiverse, the details of any holographic correspondence remain
obscure at best, so it is encouraging that the bulk-boundary
connection of AdS/CFT can be established without such knowledge.  We
will now see that a robust causal construction naturally relates bulk
points to the future boundary of the multiverse, establishing a UV-IR
connection.

\subsection{Light-cone time}
\label{sec-lc}

The goal, by analogy, is to associate to each bulk event $p$ a scale
$\epsilon(p)$ on the future boundary of the multiverse, such that
$\epsilon\to 0$ as $p$ approaches the boundary.  If the relation
$\epsilon(p)$ is known, then $\epsilon$ can be used to generate a
foliation of the bulk, where points of equal $\epsilon$ form spacelike
hypersurfaces of equal time $t(\epsilon)$, with $t\to\infty$ as
$\epsilon\to 0$.
\begin{figure*}
\begin{center}
\includegraphics[scale=.8]{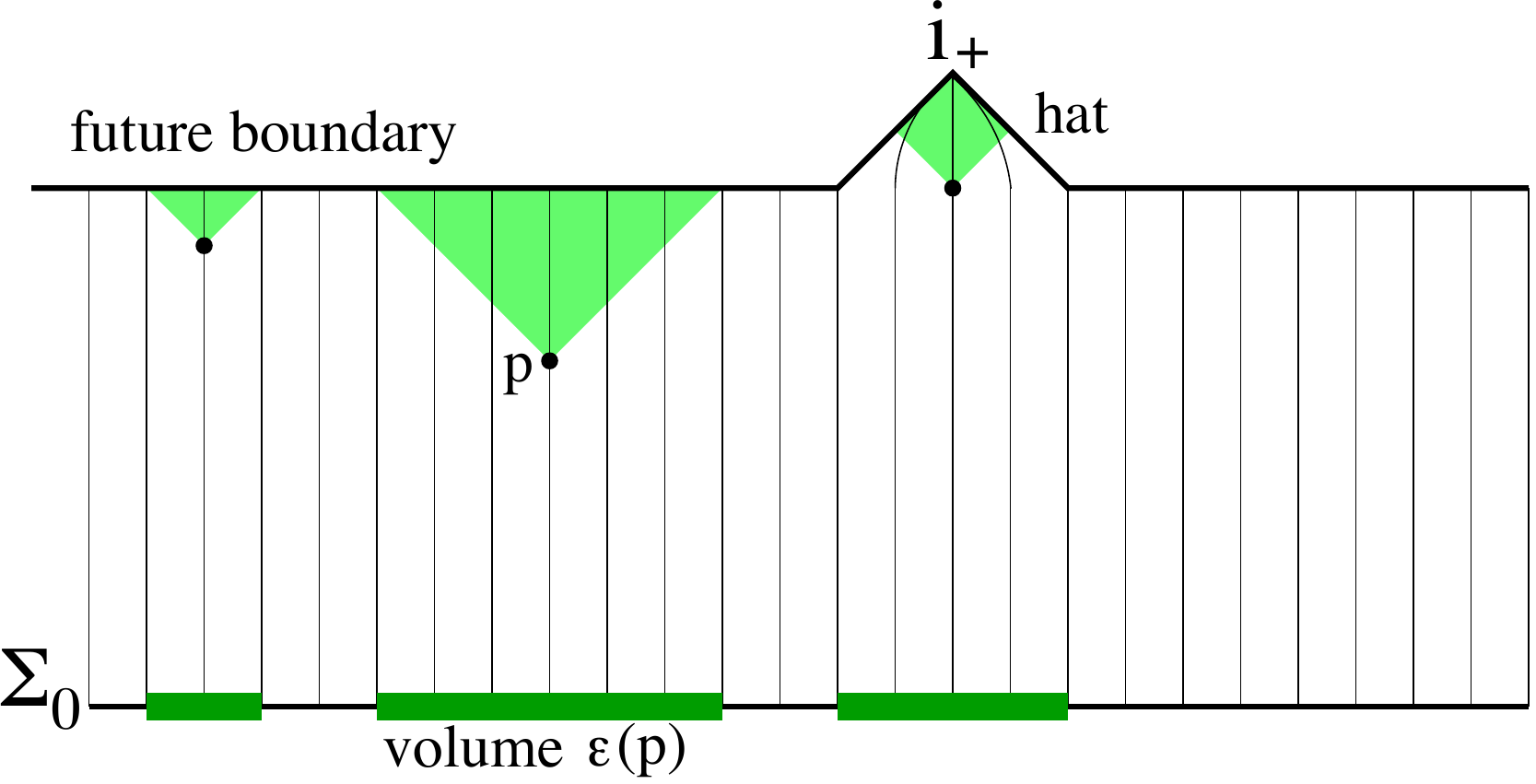}
\end{center}
\caption{The boundary scale $\epsilon$ associated with the multiverse
  event $p$ is defined as the volume (thick bars), on $\Sigma_0$, of
  those geodesics (thin vertical lines) that enter the future of $p$
  (shaded).  Other features in this diagram are discussed in
  Sec.~3.2.}
\label{fig-mversebb}
\end{figure*} 

The map $p\to\epsilon$ cannot be exactly the same as in AdS, because
the causal structure of the boundary is different.  However, it seems
appropriate to require that it be obtained, like the map $p\to\delta$
in AdS/CFT, from causality alone.  In fact, there is a very simple
construction that achieves this goal: Given a point $p$ in the
multiverse, let $\epsilon(p)$ be the volume, on the boundary, of its
chronological future $I^+(p)$ (Fig.~\ref{fig-mversebb}).

$I^+(p)$ is the set of points that can be reached from $p$ by a
future-directed timelike curve.  It can be visualized as the interior
of a future light-cone whose tip is at $p$.  To define the volume of
$I^+(p)$ on the boundary, consider a timelike geodesic congruence
$\gamma$ orthogonal to a finite spacelike hypersurface $\Sigma_0$ in
the bulk.\footnote{Here the analogy with AdS/CFT is less clear.  It
  would be nice to quantify UV scales directly on the future boundary
  of the multiverse, without reference to the congruence $\gamma$
  and/or initial hypersurface $\Sigma_0$.  Because the future boundary
  is a fractal, this task is not straightforward and will be left to
  future work.}  I will define $\epsilon$ as the volume of the
starting points, on $\Sigma_0$, of all geodesics in $\gamma$ that
eventually enter $I^+(p)$.  Note that it is not necessary to assume
that the congruence $\gamma$ be everywhere expanding.

The choice of $\Sigma_0$ is arbitrary except for the requirement that
at least one geodesic in $\gamma$ must be eternally inflating, i.e.,
must never enter a vacuum with $\Lambda\leq 0$.  The remaining freedom
in $\Sigma_0$ will not affect the UV-IR relation in the small
$\epsilon$ (i.e., late time) limit.  Therefore, it will not affect the
resulting probability measure.

The map $p\to \epsilon(p)$ defines a UV-IR relation exactly as in
AdS/CFT: Given a cut-off $\epsilon_{\rm c}$ on the boundary, remove
from the bulk all points $p$ with $\epsilon(p)<\epsilon_{\rm c}$.
Like $\delta$ in the AdS/CFT case, $\epsilon$ can be considered both a
bulk and a boundary coordinate.  Unlike $\delta$, which is a spatial
coordinate in the bulk, $\epsilon$ is a time coordinate in the bulk.
That is, the vector field $\nabla^a\epsilon$ is everywhere timelike
(except for degeneracies discussed in Sec.~\ref{sec-domains}).

To prove this, consider two points connected by a timelike curve, with
$q$ lying in the future of $p$.  Intuitively, the future light-cone of
$q$ should be nested within the light-cone of $p$ and so should be
smaller on the future boundary.  This can be made precise.  Since any
timelike curve from $q$ to a point in $I^+(q)$ can be extended by the
timelike curve connecting $p$ to $q$, $I^+(q)$ is a subset of
$I^+(p)$.  Therefore, every timelike curve that intersects with
$I^+(q)$ must also enter $I^+(p)$.  It follows that
\begin{equation} 
\epsilon(p)\geq\epsilon(q)~.
\label{eq-epeq}
\end{equation}
This result applies, in particular, as $q$ approaches $p$ along any
timelike curve, so the gradient of $\epsilon$ is everywhere timelike
and past-directed (or zero; see Sec.~\ref{sec-domains}).

Since $\epsilon$ vanishes at the future boundary and increases towards
the past, it will be convenient to define a new time coordinate $t$
that increases towards the boundary.  $t$ should depend only on
$\epsilon$, so that it will define the same spacelike hypersurfaces.
The detailed relation is irrelevant, because it affects only the rate
at which the boundary is approached.  I will choose
\begin{equation}
  t(p)=-\frac{1}{3} \log\epsilon(p) ~
\label{eq-t}
\end{equation}
so that $t\to\infty$ as $\epsilon\to 0$.  $t$ will be called
light-cone time.  Via Eq.~(\ref{eq-nt}), $t$ defines the ``light-cone
measure'' on the multiverse (Fig.~\ref{fig-lightconecut}).

\subsection{Hat domains and singular domains}
\label{sec-domains}

\begin{figure*}
\begin{center}
\includegraphics[scale=.8]{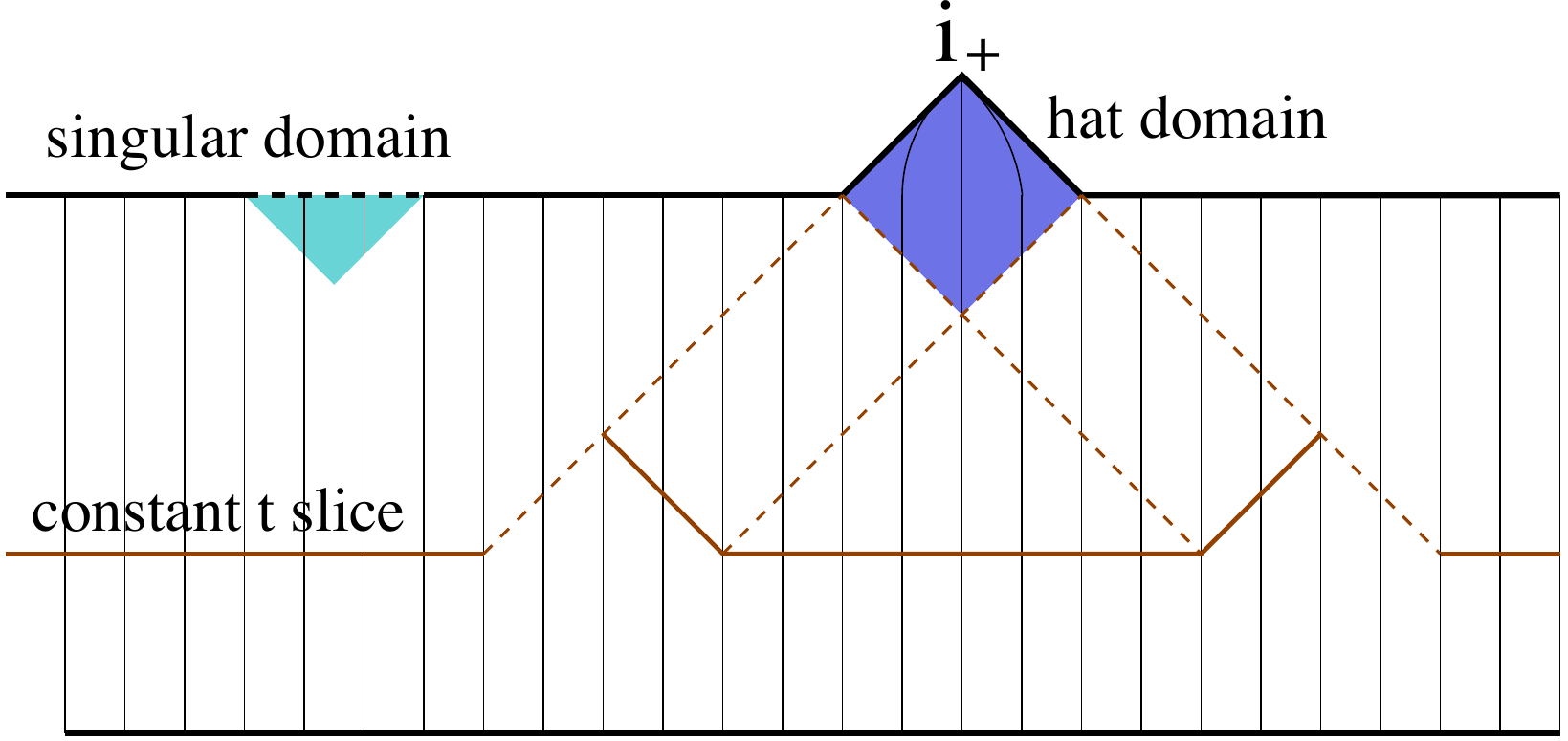}
\end{center}
\caption{Examples of singular domains and hat domains.  The light-cone
  time increases monotonically towards the future, but not strictly
  so.  Because all geodesics in the hat domain (blue spacetime region)
  end at $i_+$, the light-cone time is constant there, and has
  fractal-like bends and discontinuities in the bulk. The dashed
  diagonals are included to guide the eye.}
\label{fig-ziggyslice}
\end{figure*}

Though I have shown that $t$ increases monotonically along any
timelike curve, Eq.~(\ref{eq-epeq}) does not guarantee that it is
strictly monotonic, nor that it is continuous.  In fact, in the
presence of ``hats'', it will be neither.  Hats are the portions of
the future boundary corresponding to supersymmetric regions with
$\Lambda=0$.  Each hat is a portion of the future conformal boundary
of Minkowski space.  All geodesics in $\gamma$ that enter the past
domain of dependence of a hat, $D^-_{\rm hat}$, end up at the tip of
the hat, $i_+$ (Fig.~\ref{fig-mversebb}).  I will refer to $D^-_{\rm
  hat}$ as the ``hat domain''; an example is shown in
Fig.~\ref{fig-ziggyslice}.  For all points $p$ in a hat domain,
$I^+(p)$ contains $i_+$ and does not contain any other endpoints of
geodesics.  Therefore, $\epsilon$ (and thus, $t$) is constant in the
entire hat domain.  It is set by the volume, on $\Sigma_0$, of the
geodesics that enter the hat domain.

This has two implications.  First, a hypersurface of constant $t$,
$\Sigma_t$, will have a jagged, fractal structure, exemplified by the
solid brown line in Fig.~\ref{fig-ziggyslice}.  It will have null
portions, becoming disconnected along portions of the boundary of the
past of a hat.  This does not matter as far as Eq.~(\ref{eq-nt}) is
concerned.  The smoothness of slices is irrelevant since $N_i$ counts
events taking place in a spacetime region, defined by the set of
points $p$ occuring before the cut-off, $\{p:t(p)<t_{\rm c}\}$.

Second, and more importantly, it can lead to a divergence at finite
cut-off.  As the IR cut-off $t_{\rm c}\equiv t(\epsilon_{\rm c})$ is
increased above the value of $t$ in a particular hat, this means that
the entire spacetime four-volume of a $\Lambda=0$ bubble suddenly
becomes included below the cut-off.  If the bubble contains
observations of type $i$, then by the $SO(3,1)$ symmetry of the bubble
interior, it will contain infinitely many.  Thus, $N_i(t)$ will
diverge at finite $t$, and the cut-off will have failed to regulate
some of the fractions appearing in Eq.~(\ref{eq-nt}).

It is interesting that exactly the same divergence arises in the
causal patch measure.  This is a first hint that the equivalence shown
in Sec.~\ref{sec-equiv} below is more general than I will prove here.
This is encouraging, since the global viewpoint may admit a natural
modification that eliminates this divergence, as I will now discuss.

Among potential resolutions, perhaps the least satisfying is that the
divergence is not realized in practice, because the number of
observations is strictly zero in all hat domains.  For example,
unbroken supersymmetry may be incompatible with sufficiently complex
structures.  An infinite number of observations would still appear in
bubbles of anthropic vacua that collide with the hat domain.  However,
the domain walls between a $\Lambda=0$ vacuum, and one with suffiently
small $\Lambda$ to fit an observer, would likely have to be of the VIS
type, accelerating away from both vacua.\footnote{The last two
  sentences were pointed out to me by S.~Shenker, and by B.~Freivogel,
  respectively.}  This would imply that only microphysical slivers of
superexponentially redshifted observers are contained in the hat
domain, rendering their contribution to $N_i$ unclear.

A more interesting possibility is to modify the definition of the
volume of $I^+(p)$.  For example, one could explore definitions in
terms of the area of the boundary of $I^+(p)$ on the future boundary
of the multiverse, or in terms of the volumes of $I^-(I^+(p))$ and
$I^-(p)$ on $\Sigma_0$.  (This would have the additional advantage of
dispensing with the congruence $\gamma$.)  Hats are uniquely
distiguished on the future boundary in that they constitute its only
light-like portions.  Thus, it is quite plausible that there exists a
definition of $\epsilon$ which reduces to the one I have given, except
in hat domains.

A third possibility is that the direct analogy with AdS/CFT breaks
down in hat domains.  Then the measure of Eq.~(\ref{eq-nt}), with the
cut-off defined by Eq.~(\ref{eq-t}), should simply not be applied to
events in hat regions.  In fact, it is plausible that it may break
down also in ``singular domains'', defined as the past domains of
dependence of singular portions of the future boundary.  This would
exclude from consideration all bubbles with negative cosmological
constant, and the interior of black holes.  The remaining set
$M-D^-(\mbox{all hats})-D^-(\mbox{all future singularities})$, where
the cut-off does apply, is the ``eternal domain'', consisting of the
set of points whose future includes at least one eternally inflating
geodesic.  This includes most events taking place in metastable
de~Sitter vacua (such as, presumably, ours).

For the remainder of this paper, I will follow Garriga and Vilenkin in
adopting this last, most conservative viewpoint.  Unlike GV, I will
not attempt to fill the resulting gaps and formulate a separate cut-off
for hat domains and singular domains.  Eqs.~(\ref{eq-nt}) and
(\ref{eq-t}) suffice to compute the relative probabilities of events
that occur in the eternal domain, which is all I will need in
Sec.~\ref{sec-equiv} below.  However, in order to develop the
light-cone measure further it will clearly be important to explore
alternative formulations in hat domains, and possibly also in singular
domains.  I leave this to future work.

\section{The Garriga-Vilenkin parameter $\lambda$ 
and the scale factor  cut-off}
\label{sec-gv}

The light-cone construction of the previous section is a concrete
realization of the general idea of a boundary cut-off for the
multiverse~\cite{GarVil08}.  However, it differs from the
bulk-boundary connection that Garriga and Vilenkin themselves have
outlined.  The GV proposal does not involve future light-cones; in
fact, the future boundary plays no role at all in assigning a scale to
each bulk point.  GV introduce a physical length scale $\lambda$,
which depends on the physical process whose probability one would like
to compute, and which is defined loosely as the ``resolution''
required to ``describe'' or identify this process.

The parameter $\lambda$ has no analogue in the AdS/CFT correspondence.
Moreover, its definition remains too vague for a meaningful comparison
of the GV proposal to the light-cone proposal.\footnote{Of course, the
  {\em fact\/} that it is more sharply defined favors the light-cone
  proposal. So do its economy, its greater generality (it does not
  require $\gamma$ to be everywhere expanding), and its AdS/CFT
  pedigree.  Moreover, under at least one plausible interpretation of
  ``resolution'', the proposed definition of $\lambda$ contradicts
  observation.  For example, consider the probability distribution of
  the frequency of CMB photons.  To ``resolve'' a given photon,
  $\lambda$ should be chosen of order its wavelength.  By
  Eq.~(\ref{eq-gvsf}), the detection of high energy photons would then
  be suppressed relative to the black body curve, in proportion to the
  third power of the wavelength.  (GV consider the possibility that on
  subhorizon scales, $\lambda$ may not be related to the scales one
  would naively associate with various phenomena.  But this further
  obscures its definition precisely for those phenomena we can
  actually measure.  And it may not help: In our universe, the entire
  CMB spectrum will be stretched to superhorizon scales.)}  However,
it is convenient to treat $\lambda$ as a free parameter that may or
may not dependent on the spacetime point $p$.  In this section, I will
establish some properties of this ``generalized'' GV construction.
These results will be useful in Sec.~\ref{sec-equiv}, where I will
show that the light-cone construction {\em defines\/} a particular
choice of $\lambda$, and that this choice reproduces the causal patch
measure.

In Sec.~\ref{sec-gv1}, I will review the GV construction.  For fixed
$\lambda$, it reproduces the scale factor time in the
bulk~\cite{GarVil08} (Sec.~\ref{sec-gv2}).  In Sec.~\ref{sec-gv3}, I
will generalize this result.  I will consider the GV relation in the
case where $\lambda$ does depend on the type of bulk event in
question.  I find the resulting class of measures to be simply related
to the scale factor measure: The relative probability $p_i/p_j$
differs by $\lambda_i^3/\lambda_j^3$ from the prediction of the scale
factor measure.

\subsection{The Garriga-Vilenkin bulk-to-boundary relation}
\label{sec-gv1}

Again, one begins with the (arbitrary) choice of a finite initial
hypersurface, $\Sigma_0$ and its orthogonal congruence of timelike
geodesics, $\gamma$, at least one member of which must be eternal.
Instead of the future light-cone at $p$, GV consider a physical length
scale $\lambda(p)$, which is supposed to be chosen somewhat smaller
than the characteristic scale of the physical process at $p$ whose
probability one would like to compute.
\begin{figure*}
\includegraphics[scale=.8]{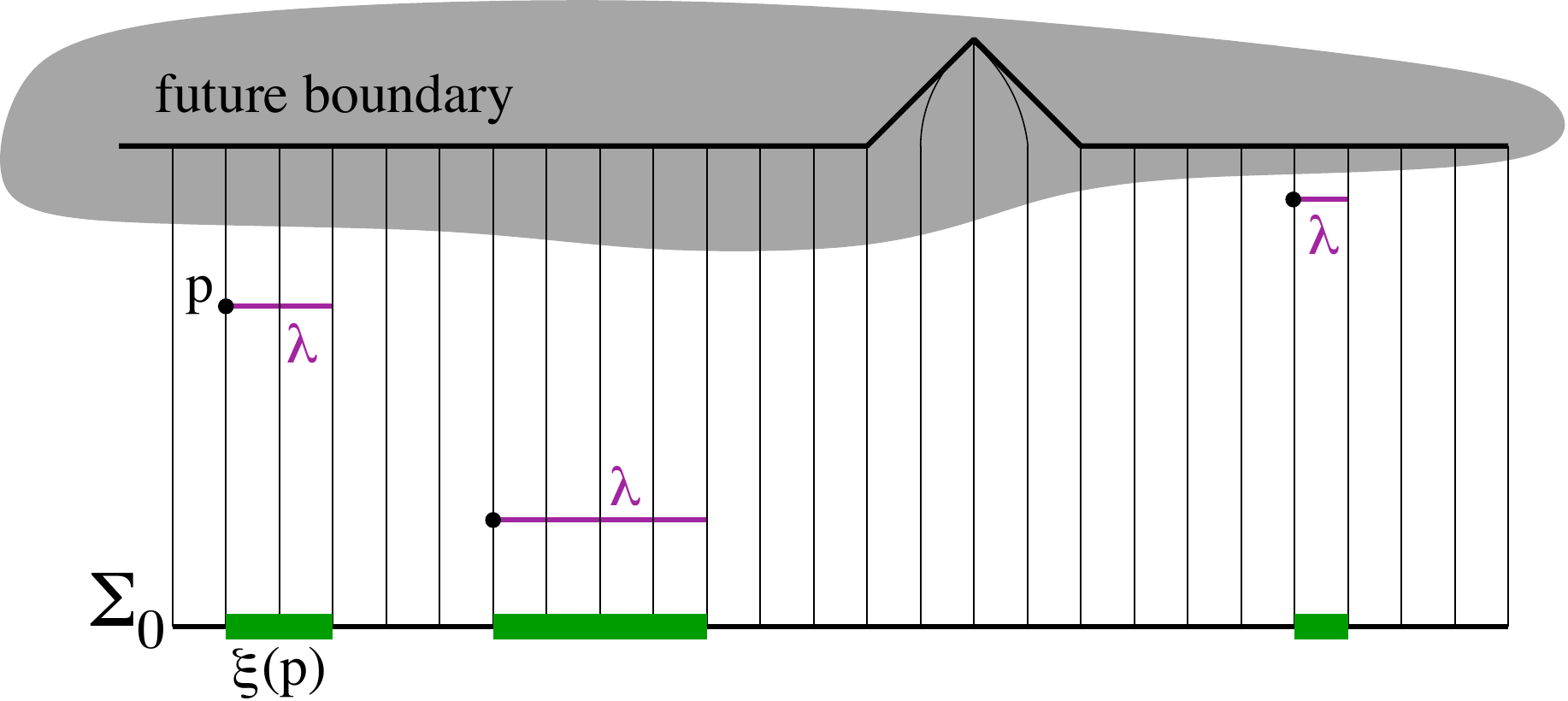}
\caption{Garriga and Vilenkin consider a physical length scale
  $\lambda$ at the multiverse event $p$ (purple intervals).  This
  length is transported along the congruence $\gamma$ (thin vertical
  lines) back to the initial surface $\Sigma_0$, thereby contracting
  to physical size $\xi$ (thick green bars).  This construction of
  $\xi(p)$ does not actually involve the future boundary or asymptotic
  regime (shaded), though one can choose to picture $\xi$ as a
  boundary scale by transporting the interval to the future boundary
  along $\gamma$.  In the example shown, $\lambda$ is chosen the same
  at all points, so $\xi$ will define slices of constant scale
  factor.}
\label{fig-gvtime}
\end{figure*} 

To associate a boundary length scale $\xi$ to the bulk event $p$, one
transports a spatial interval of length $\lambda$ from $p$ to the
future boundary along the congruence $\gamma$, allowing it to expand
with the congruence along the way.  GV define $\xi$ as the projection,
onto $\Sigma_0$, of the resulting boundary interval, again along the
congruence $\gamma$.  Viewed as a bulk coordinate, $\xi$ is
past-directed, so it is convenient to define the time coordinate
\begin{equation}
  \hat\eta=-\log\xi~,
\end{equation}
which satisfies $\hat\eta\to\infty$ as $\xi\to 0$~.  A given UV cut-off
$\xi_{\rm c}$ on the boundary thus defines a late time cut-off
$\hat\eta_{\rm c}$ in the bulk.  Via Eq.~(\ref{eq-nt}), the limit
$\xi_{\rm c}\to 0$ defines a measure in the multiverse.

In order for $\hat\eta$ to increase monotonically as $\xi$ is
decreased, it is necessary that the congruence $\gamma$ be everywhere
expanding.  This condition is restrictive; it is violated, for
example, in gravitationally collapsed regions such as our galaxy.
Note that it was not needed for defining the light-cone time $t$ in
the previous section.

One could have obtained $\xi(p)$ and $\hat\eta(p)$ more directly by
transporting the physical interval $\lambda$ from $p$ back in time to
the initial hypersurface $\Sigma_0$, without first making an excursion
forward in time to the boundary (Fig.~\ref{fig-gvtime}). Thus, in the
end, no properties of the boundary play any role in the construction.
This contrasts with the light-cone time, which cannot be defined
without knowledge of the asymptotic future.  I will work with this
simpler, more direct definition, because it makes it easy to see that
$\hat\eta$ is closely related to the scale factor time.

\subsection{Constant $\lambda$ and the scale factor time}
\label{sec-gv2}

The scale factor $a$ at a point $p$ in the congruence $\gamma$ is
defined by
\begin{equation}
a(p)^3=\frac{d V_p}{d V_{\Sigma_0}}~.
\end{equation}
Here, $d V$ is the physical volume element spanned by the geodesic
that includes $p$ and its infinitesimally neighboring geodesics in the
congruence. Thus, the local scale factor $a$ measures how much this
physical volume has expanded along the geodesic connecting $p$ to the
initial surface $\Sigma_0$.  The scale factor time is defined by
\begin{equation}
\eta\equiv \log a~.
\label{eq-eta}
\end{equation}

Let us assume isotropic expansion and relax the requirement that $d V$
be infinitesimal.  Then it follows from the above definitions of $\xi$
and $a$ that
\begin{equation}
  a(p)=\frac{\lambda}{\xi}~,  
\label{eq-alxi}
\end{equation}
so $\hat\eta$ is related to the scale factor time by a
$\lambda$-dependent shift:
\begin{equation}
\hat\eta(p)=\eta(p)-\log \lambda(p)~.
\end{equation}
If $\lambda$ is constant, then $\eta$ and $\hat\eta$ define the same
hypersurfaces and therefore the same probability measure, the scale
factor measure.  

\subsection{General $\lambda$}
\label{sec-gv3}

Suppose now that $\lambda$ is not constant, but that every type of
event $i$ whose probability we would like to compute is associated to
a unique value of $\lambda$, $\lambda_i$.  (For example, this would be
the case if, as Garriga and Vilenkin suggest, $\lambda$ is related to
the bulk resolution required to describe the physics of $i$.
Moreover, I will show below that the light-cone cut-off is recovered,
$\hat\eta=t$, for some choice of $\lambda_i$.)  This means that in
effect, one is using different scale factor times, shifted by
$\log\lambda_i$, to define the number of such events below the cut-off,
$\hat N^{\rm GV}_i(\hat\eta)$:
\begin{equation}
  N^{\rm GV}_i(\hat\eta)=N^{\rm SF}_i(\eta_i)~,
\end{equation}
where
\begin{equation}
\eta_i\equiv \hat\eta+\log\lambda(p)~.
\end{equation}

The scale factor cut-off exhibits attractor behavior.  At late times,
the number of events $i$ grows exponentially,
\begin{equation}
  \lim_{\eta\to\infty} 
  \frac{N^{\rm SF}_i(\eta+\Delta \eta)}{N^{\rm SF}_i(\eta)} =
  e^{(3-q)\Delta \eta}~,
\end{equation}
where $q\ll 1$ in a realistic landscape of vacua.  This universal
behavior was found in Ref.~\cite{GarSch05} for the special case where
$i$ is the observation of a particular metastable vacuum in the
landscape.  It holds regardless of how the event $i$ is defined.  For
example, $i$ could refer to the observation of a particular CMB
temperature.  The only exception are terminal states, whose volume
fraction grows as $1-e^{-q\eta}$ at late times (this defines $q$), but
which are not observed in any case.\footnote{Terminal vacua can
  contain observations.  I am assuming only that they will cease to do
  so after some finite time.  This is clearly the case for regions
  with negative cosmological constant (which, however, cannot be
  described by an everywhere-expanding congruence), and it may also be
  the case for regions with zero cosmological constant.  A terminal
  state is one in whose future no observations of any type occur.}

The previous three equations imply that
\begin{equation}
  \lim_{\hat\eta\to\infty} 
  \frac{N^{\rm GV}_i(\hat\eta)}{N^{\rm SF}_i(\hat\eta)}= 
  \lambda_i^{3-q}~.
\end{equation}
By Eq.~(\ref{eq-nt}), it follows that the relative probabilities
defined by the GV cut-off, and those computed from the SF cut-off, are
simply related by powers of $\lambda_i$:
\begin{equation} 
\left(\frac{p_i}{p_j}\right)_{\rm GV}=
\left(\frac{p_i}{p_j}\right)_{\rm SF}
\left(\frac{\lambda_i}{\lambda_j}\right)^{3-q}~.
\label{eq-gvsf}
\end{equation}
For $q$ sufficiently small, $1-(\lambda_i/\lambda_j)^q\ll 1$ for all
pairs $i,j$, and one can set $q\to 0$ in the above formula.

\section{Equivalence between light-cone and causal patch measures}
\label{sec-equiv}

In this section, I will consider the probability measure defined by
the light-cone cut-off of Sec.~\ref{sec-multi}.  I will show that in a
broad regime of bulk regions, the resulting measure is equivalent to
the causal patch measure, with initial conditions dictated by the
attractor behavior of the light-cone slicing.  This relation is
remarkable, since the causal patch measure was formulated as a
``local'' measure without reference to a global geometry.

The argument will proceed by formulating the light-cone cut-off as a
special case of the GV cut-off corresponding to a particular
spacetime-dependent choice of $\lambda(p)$.  (This is purely for
convenience; in particular, this choice of $\lambda$ does not appear
to meet the criteria Garriga and Vilenkin have described: As we shall
see, the light-cone does not relate $\lambda$ to the scale of
experiments that may be taking place at $p$, nor does it necessarily
relate $\lambda$ to the Hubble radius at $p$.)  The advantage of this
viewpoint is that the dependence of $\lambda$ on $p$ quantifies how
the light-cone cut-off differs from the scale factor cut-off, which
corresponds to constant $\lambda$.  I will show that this {\em
  difference\/} is the same as the (known) difference between the
causal patch measure and the scale factor measure.  Therefore, the
light-cone cut-off must be equivalent to the causal patch measure.

My analysis will not be completely general.\footnote{See, however, the
  Note added below.}  Because I will use the scale factor cut-off as an
intermediary, I will need to assume that the congruence $\gamma$ is
everywhere expanding.  (Neither the causal patch cut-off nor the
light-cone cut-off require this assumption.)  Moreover, for simplicity,
I will only consider observations that take place in approximately
homogeneous, isotropic regions with positive cosmological constant and
negligible spatial curvature (such as ours).  In such regions, the
congruence $\gamma$ will be comoving with the homogeneous matter or
radiation.  Moreover, over distances less than the curvature scale,
hypersurfaces of constant scale factor time will have constant
density~\cite{BouFre08b}, and will thus agree with the usual
Friedmann-Robertson-Walker slices.
\begin{figure*}
\begin{center}
\includegraphics[scale=1]{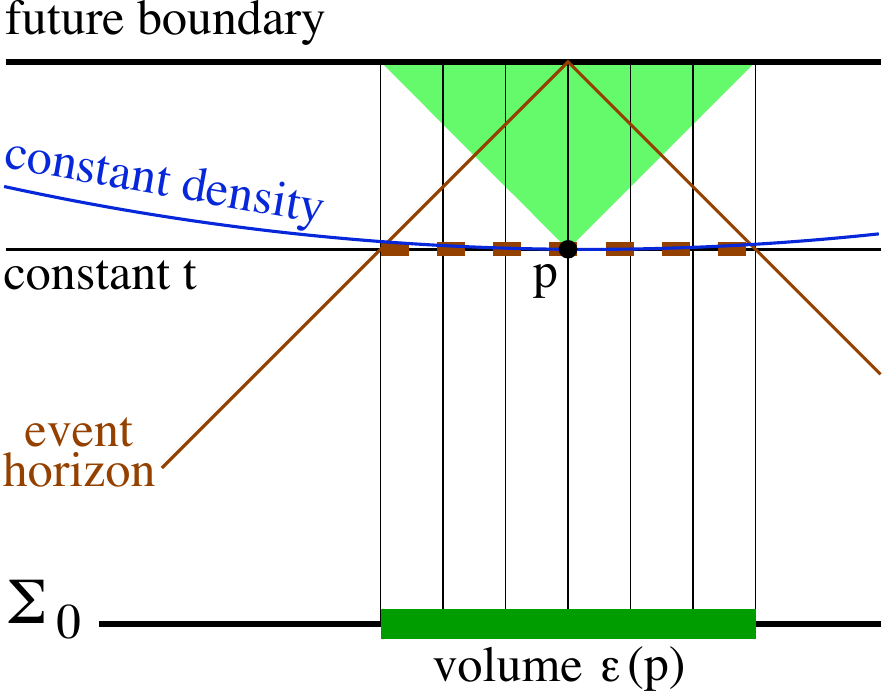}
\end{center}
\caption{Light-cone time as a special case of the generalized
  Garriga-Vilenkin construction. In homogeneous, isotropic, spatially
  flat regions, the future light-cone of $p$ (shaded) and the event
  horizon at $p$ have the same comoving volume.  At $p$, the
  corresponding physical volume is $V_{\rm EH}$ (thick dashed).  Thus,
  by choosing the GV parameter to be $\lambda=V_{\rm EH}^{1/3}$, one
  recovers the light-cone time slicing.  Combined with earlier
  results, this implies that the light-cone measure is equivalent to
  the causal diamond measure.}
\label{fig-eventhorizon}
\end{figure*} 

The boundary scale $\epsilon(p)$ is the volume, on $\Sigma_0$, of the
geodesics entering the chronological future $I^+(p)$.  By homogeneity
and isotropy, a geodesic on the constant density hypersurface
containing $p$ will enter $I^+(p)$ if and only if it is inside the
event horizon of the geodesic passing through $p$ (see
Fig.~\ref{fig-eventhorizon}).  Because curvature is negligible, this
constant density hypersurface agrees with the hypersurface defined by
the scale factor at $p$, $a=a_p$, at least on the scale of the event
horizon.  Thus, $\epsilon(p)$ can be obtained by scaling the physical
volume of the event horizon at $p$ back to $\Sigma_0$:
\begin{equation}
\epsilon(p)=a(p)^{-3} V^{\rm EH}(p)
\end{equation}
In order to obtain the same result from the GV prescription (i.e., in
order for $\epsilon(p)=\xi(p)^3$), one would need to choose
\begin{equation}
  \lambda(p)^3=V^{\rm EH}(p)~,
\label{eq-lv}
\end{equation}
by Eq.~(\ref{eq-alxi}).\footnote{The light-cone defines a volume on
  the boundary, whereas $\lambda$ defines a length scale.  In fact,
  however, both prescriptions use a congruence of timelike geodesics
  to quantify scales on the boundary, and therefore both fundamentally
  define a volume.  The expansion along a congruence is generically
  anisotropic, so the length $\xi$ of the projection of the bulk scale
  $\lambda$ onto the initial surface $\Sigma_0$ depends on its spatial
  orientation.  To avoid this ambiguity, $\lambda$ and $\xi$ should be
  understood as third roots of volumes, say of cubes of linear size
  $\lambda$ at $p$.  It is interesting that the holographic
  construction of a measure does not seem to require a metric
  structure (definition of lengths) on the boundary, but only a notion
  of volumes, which is weaker.}  Whereas with fixed $\lambda$, the GV
prescription reproduces the scale factor cut-off, we now see that with
$\lambda$ as in Eq.~(\ref{eq-lv}), it reproduces the light-cone
cut-off.

Now consider the relative probability of events of type $i$ and $j$.
Without loss of generality I will assume that $i$ has been
sufficiently narrowly specified that the physical volume of the event
horizon is the same, $V^{\rm EH}_i$, at all points $p$ at which $i$
takes place; similarly for $j$.  (For example, since $V^{\rm EH}_i$ is
observable, this can be accomplished by including it in the
specification of $i$.)  By Eqs.~(\ref{eq-gvsf}) and (\ref{eq-lv}), the
relative probability of $i$ and $j$ computed from the light-cone
measure is related to that computed from the scalefactor measure by
the ratio of event horizon volumes:
\begin{equation}
  \left(\frac{p_i}{p_j}\right)_{\rm LC}=
  \left(\frac{p_i}{p_j}\right)_{\rm SF}\frac{V^{\rm EH}_i}{V^{\rm EH}_j}~,
\label{eq-lcsf}
\end{equation}
where I have used $q\ll 1$.

I will now show that the right hand side is the relative probability
defined by the causal patch measure.  In Ref.~\cite{BouFre08b}, the
scale factor measure was reformulated in terms of the evolution of a
single geodesic of infinitesimal thickness $dV$ (the dark shaded
region in Fig.~\ref{fig-causalcut}):
\begin{equation}
  \left(\frac{p_i}{p_j}\right)_{\rm SF}=
  \frac{\langle dN_i/dV\rangle}{\langle dN_j/dV\rangle}
\label{eq-sflocal}
\end{equation}
where $dN_i$ are the number of events of type $i$ occuring in the
worldtube of thickness $dV$.  The $\langle\rangle$ signs indicate that
one must average over possible histories, to account for the fact that
a geodesic starting with given initial conditions has finite branching
ratios for transitions into different vacua.  For this ``local''
formulation to be equivalent to the scale factor measure as defined in
Eqs.~(\ref{eq-nt}) and (\ref{eq-eta}), the initial conditions for the
geodesic must reflect the attractor behavior of the scale factor
slicing.  That is, another average is taken in which geodesics
starting in the state $i$ receive weight $V_i$, where $V_i$ is the
volume fraction occupied by $i$ on constant scale factor hypersurfaces
at late times.  As shown in Ref.~\cite{SchVil06}, the nonterminal
volume distribution in the attractor regime is dominated by the
metastable vacuum $*$ with the slowest decay rate,
\begin{equation}
  \tilde V_*\gg \tilde V_i~~~\mbox{for~all}~i\neq *~.
\label{eq-dom}
\end{equation}
Thus, to excellent approximation, the geodesic can be taken to start
out in the dominant vacuum, $*$.

The causal patch measure~\cite{Bou06} is similar to the ``local''
formulation of the scale factor measure I have just reviewed, with two
differences: First, the ``thickness'' of the geodesic is not fixed,
nor is it infinitesimal.  Instead, one includes events in the entire
causal past of its future endpoint, i.e., in the interior of its event
horizon (the entire shaded region in
Fig.~\ref{fig-causalcut}).\footnote{One can restrict the causal patch
  further by including only points which are {\em also\/} contained in
  the future of the geodesic's starting point.  In a causal diagram
  this gives the surviving four-volume a diamond shape.  However, this
  additional restriction tends to affect only the earliest portions of
  the four-volume and is irrelevant here.}
\begin{equation}
  \left(\frac{p_i}{p_j}\right)_{\rm CD}=
  \frac{\langle N_i^{\rm EH} \rangle}{\langle N_j^{\rm EH} \rangle}
\label{eq-cd}
\end{equation}
Secondly, initial conditions were not specified in Ref.~\cite{Bou06},
but were left to a future theory of initial conditions.

Combining Eqs.~(\ref{eq-lcsf}) and (\ref{eq-sflocal}), it follows that
\begin{equation}
  \left(\frac{p_i}{p_j}\right)_{\rm LC}=
  \frac{\langle dN_i/dV\rangle}{\langle dN_j/dV\rangle}
  \frac{V^{\rm EH}_i}{V^{\rm EH}_j}~.
\end{equation}
Assuming homogeneity, $\frac{dN_i}{dV} V^{\rm EH}_i$ is just the
number $N^{EH}_i$ of events of type $i$ that occur within the event
horizon of the geodesic.  Therefore, by Eq.~(\ref{eq-cd}), the
light-cone cut-off is equivalent to the causal patch measure:
\begin{equation}
  \left(\frac{p_i}{p_j}\right)_{\rm LC}=
  \left(\frac{p_i}{p_j}\right)_{\rm CD}~.
\end{equation}
At the level of approximation I have used, the initial conditions for
the geodesic spanning the causal patch are determined by the attractor
regime of the scale factor measure.  Strictly speaking, they are
determined by the attractor regime defined by hypersurfaces of equal
light-cone time $t$, which will be slightly different for small but
finite $q$.  Either way, in a realistic landscape, it is an excellent
approximation to start the geodesic in the longest lived metastable
vacuum, $*$, and to average over possible decoherent
histories.\footnote{This would modify the analysis of
  Ref.~\cite{BouYan07}, where Planck scale initial conditions were
  assumed.}

\section{Discussion}
\label{sec-discussion}

That two measures are equivalent does not mean that they are right.
Ref.~\cite{BouFre08b} showed that the scale factor measure admits both
a global and a local formulation.  Eq.~(\ref{eq-gvsf}) generalizes
this global-local duality to an infinite family of pairs of equivalent
measures, parameterized by different choices of $\lambda$.  Yet, the
pair we have singled out, the causal patch measure and the light-cone
time cut-off, occupy a special place in this family.  Both arose from
specific (but different) approaches related to holography.  Each of
these two approaches was developed independently of the other, and
independently of the duality that has now been shown to relate them.

In fact, three different roads have led us to the same place: (1) The
causal patch measure~\cite{Bou06} was motivated by black hole
complementarity, well before (2) its numerous phenomenological
successes were understood, such as the absence of Boltzmann
brains~\cite{BouFre06b}, Boltzmann babies~\cite{BouFre07}, a
``Q-catastrophe''~\cite{Bou06}, and of certain runaway
problems~\cite{BouHar07,CliFre07}; and its excellent agreement with
the observed value of the cosmological constant~\cite{BouHar07}.  We
can now add to this a new piece of evidence: (3) The global
``holographic'' cut-off advocated by Garriga and
Vilenkin~\cite{GarVil08}, and motivated by the UV-IR connection of
AdS/CFT, reproduces the relative probabilities obtained from the
causal patch measure.

The causal patch and global approaches complement each other.  The
restriction to a causal patch avoids the apparent cloning of quantum
information by Hawking radiation, which plagues the global description
of any spacetime with evaporating black holes, such as the multiverse.
But this is a very subtle quantum effect, and it need not preclude the
usefulness of the global picture for other purposes.  For example, the
causal patch can be constructed with any initial conditions.  Although
it is perfectly conceivable that an independent theory determines
initial conditions, the causal patch measure has been criticized for
being less predictive, on this count, than global measures whose
predictions are largely independent of initial conditions.  Indeed, it
now appears that the initial conditions for the causal patch are
completely determined by the duality with the global picture.  For the
duality to hold, one must weight all possible initial states in the
causal patch in proportion to their volume fraction in the attractor
regime of the light-cone time slicing.

The multiverse complementarity uncovered here is not like black hole
complementarity, if the latter is interpreted as the statement that
complete information about the physics behind an observer's horizon is
available in his causal patch in scrambled form.  (This property, in
any case, seems peculiar to black holes arising from scattering
experiments in otherwise empty space, and would not appear to
generalize to cosmology.  For example, the interior of a small black
hole cannot contain information about a large distant galaxy,
scrambled or not.)  In the multiverse, neither the global nor the
causal patch picture contains the whole truth.  Perhaps a holographic
theory on the future boundary will unify the global and causal patch
descriptions of the bulk, allowing each to emerge in a suitable limit.

Many questions remain open. How should hat domains be treated?  It may
be possible to eliminate the divergences discussed in
Sec.~\ref{sec-domains} by a modification that affects the measure only
in hats.  This may reveal a connection to the approach of Freivogel
{\em et al.}~\cite{FreSek06}, who have focussed on the two-dimensional
rim of the hat; or it might suggest that the null portions of the
future boundary play a nontrivial role.  How should singular domains
be treated?  Here the situation is, in a sense, reversed: Light-cone
time is well-defined, but it is totally unclear how to approach the
formulation of a quantum gravity theory on the boundary.  It seems
unlikely to me that this problem is any simpler for big crunch
singularities than it is for the description of the interior of a
Schwarzschild black hole.  Perhaps light-cone time will yield a fresh
perspective on the problem of spacelike singularities.

\paragraph{Note added} The equivalence between the causal patch and
light-cone cut-offs holds independently of the assumptions made in
Sec.~\ref{sec-equiv} \cite{BouYanTA}.

\acknowledgments I thank B.~Freivogel and I.~Yang for very helpful
discussions.  This work was supported by the Berkeley Center for
Theoretical Physics, by a CAREER grant (award number 0349351) of the
National Science Foundation, and by the US Department of Energy under
Contract DE-AC02-05CH11231.

\bibliographystyle{utcaps}
\bibliography{all}

\providecommand{\href}[2]{#2}\begingroup\raggedright\begin{thebibliography}{10}

\bibitem{GarVil08}
J.~Garriga and A.~Vilenkin, ``{Holographic Multiverse},''
  \href{http://dx.doi.org/10.1088/1475-7516/2009/01/021}{{\em JCAP} {\bf 0901}
  (2009)  021},
\href{http://arxiv.org/abs/0809.4257}{{\tt arXiv:0809.4257 [hep-th]}}.

\bibitem{Mal97}
J.~Maldacena, ``The Large {$N$} limit of superconformal field theories and
  supergravity,'' {\em Adv. Theor. Math. Phys.} {\bf 2} (1998)  231,
  \href{http://arxiv.org/abs/hep-th/9711200}{{\tt hep-th/9711200}}.

\bibitem{SusTho93}
L.~Susskind, L.~Thorlacius, and J.~Uglum, ``The Stretched horizon and black
  hole complementarity,'' {\em Phys. Rev. D} {\bf 48} (1993)  3743,
\href{http://arxiv.org/abs/hep-th/9306069}{{\tt hep-th/9306069}}.

\bibitem{Bou06}
R.~Bousso, ``Holographic probabilities in eternal inflation,'' {\em Phys. Rev.
  Lett.} {\bf 97} (2006)  191302,
\href{http://arxiv.org/abs/hep-th/0605263}{{\tt hep-th/0605263}}.

\bibitem{BouFre06a}
R.~Bousso, B.~Freivogel, and I.-S. Yang, ``Eternal inflation: The inside
  story,'' {\em Phys. Rev. D} {\bf 74} (2006)  103516,
\href{http://arxiv.org/abs/hep-th/0606114}{{\tt hep-th/0606114}}.

\bibitem{BouYan07}
R.~Bousso and I.-S. Yang, ``Landscape Predictions from Cosmological Vacuum
  Selection,'' {\em Phys. Rev. D} {\bf 75} (2007)  123520,
\href{http://arxiv.org/abs/hep-th/0703206}{{\tt hep-th/0703206}}.

\bibitem{BouHar07}
R.~Bousso, R.~Harnik, G.~D. Kribs, and G.~Perez, ``Predicting the cosmological
  constant from the causal entropic principle,'' {\em Phys. Rev. D} {\bf 76}
  (2007)  043513,
\href{http://arxiv.org/abs/hep-th/0702115}{{\tt hep-th/0702115}}.

\bibitem{SusWit98}
L.~Susskind and E.~Witten, ``The holographic bound in {A}nti-de~{S}itter
  space.'' 1998.

\bibitem{DGSV08}
A.~De~Simone, A.~H. Guth, M.~P. Salem, and A.~Vilenkin, ``{Predicting the
  cosmological constant with the scale-factor cutoff measure},''
\href{http://arxiv.org/abs/0805.2173}{{\tt arXiv:0805.2173 [hep-th]}}.

\bibitem{LinLin94}
A.~Linde, D.~Linde, and A.~Mezhlumian, ``From the Big Bang theory to the theory
  of a stationary universe,'' {\em Phys. Rev. D} {\bf 49} (1994)  1783--1826,
  \href{http://arxiv.org/abs/gr-qc/9306035}{{\tt gr-qc/9306035}}.

\bibitem{GarLin94}
J.~Garc{\'{\i}}a-Bellido, A.~Linde, and D.~Linde, ``Fluctuations of the
  gravitational constant in the inflationary {B}rans-{D}icke cosmology,'' {\em
  Phys. Rev. D} {\bf 50} (1994)  730--750,
  \href{http://arxiv.org/abs/astro-ph/9312039}{{\tt astro-ph/9312039}}.

\bibitem{GarLin94a}
J.~Garc{\'{\i}}a-Bellido and A.~D. Linde, ``Stationarity of inflation and
  predictions of quantum cosmology,'' {\em Phys. Rev.} {\bf D51} (1995)
  429--443, \href{http://arxiv.org/abs/hep-th/9408023}{{\tt hep-th/9408023}}.

\bibitem{GarLin95}
J.~Garc{\'{\i}}a-Bellido and A.~Linde, ``Stationary solutions in
  {B}rans-{D}icke stochastic inflationary cosmology,'' {\em Phys. Rev. D} {\bf
  52} (1995)  6730--6738, \href{http://arxiv.org/abs/gr-qc/9504022}{{\tt
  gr-qc/9504022}}.

\bibitem{Lin06}
A.~Linde, ``Sinks in the Landscape, {B}oltzmann {B}rains, and the Cosmological
  Constant Problem,'' {\em JCAP} {\bf 0701} (2007)  022,
\href{http://arxiv.org/abs/hep-th/0611043}{{\tt hep-th/0611043}}.

\bibitem{BouRan01}
R.~Bousso and L.~Randall, ``Holographic domains of {A}nti-de {S}itter space,''
  {\em JHEP} {\bf 04} (2002)  057,
\href{http://arXiv.org/abs/hep-th/0112080}{{\tt hep-th/0112080}}.

\bibitem{CEB2}
R.~Bousso, ``Holography in general space-times,'' {\em JHEP} {\bf 06} (1999)
  028,
\href{http://arxiv.org/abs/hep-th/9906022}{{\tt hep-th/9906022}}.

\bibitem{CEB1}
R.~Bousso, ``A covariant entropy conjecture,'' {\em JHEP} {\bf 07} (1999)  004,
\href{http://arxiv.org/abs/hep-th/9905177}{{\tt hep-th/9905177}}.

\bibitem{Bou02}
R.~Bousso, ``The holographic principle,'' {\em Rev. Mod. Phys.} {\bf 74} (2002)
   825,
\href{http://arXiv.org/abs/hep-th/0203101}{{\tt hep-th/0203101}}.

\bibitem{GarSch05}
J.~Garriga, D.~Schwartz-Perlov, A.~Vilenkin, and S.~Winitzki, ``Probabilities
  in the inflationary multiverse,'' {\em JCAP} {\bf 0601} (2006)  017,
\href{http://arxiv.org/abs/hep-th/0509184}{{\tt hep-th/0509184}}.

\bibitem{BouFre08b}
R.~Bousso, B.~Freivogel, and I.-S. Yang, ``{Properties of the scale factor
  measure},''
\href{http://arxiv.org/abs/0808.3770}{{\tt arXiv:0808.3770 [hep-th]}}.

\bibitem{SchVil06}
D.~Schwartz-Perlov and A.~Vilenkin, ``Probabilities in the
  {B}ousso-{P}olchinski multiverse,'' {\em JCAP} {\bf 0606} (2006)  010,
\href{http://arxiv.org/abs/hep-th/0601162}{{\tt hep-th/0601162}}.

\bibitem{BouFre06b}
R.~Bousso and B.~Freivogel, ``A paradox in the global description of the
  multiverse,'' {\em JHEP} {\bf 06} (2007)  018,
\href{http://arxiv.org/abs/hep-th/0610132}{{\tt hep-th/0610132}}.

\bibitem{BouFre07}
R.~Bousso, B.~Freivogel, and I.-S. Yang, ``{Boltzmann babies in the proper time
  measure},'' \href{http://dx.doi.org/10.1103/PhysRevD.77.103514}{{\em Phys.
  Rev.} {\bf D77} (2008)  103514},
\href{http://arxiv.org/abs/0712.3324}{{\tt arXiv:0712.3324 [hep-th]}}.

\bibitem{CliFre07}
J.~M. Cline, A.~R. Frey, and G.~Holder, ``Predictions of the causal entropic
  principle for environmental conditions of the universe,''
\href{http://arxiv.org/abs/arXiv:0709.4443 [hep-th]}{{\tt arXiv:0709.4443
  [hep-th]}}.

\bibitem{FreSek06}
B.~Freivogel, Y.~Sekino, L.~Susskind, and C.-P. Yeh, ``A holographic framework
  for eternal inflation,'' {\em Phys. Rev. D} {\bf 74} (2006)  086003,
\href{http://arxiv.org/abs/hep-th/0606204}{{\tt hep-th/0606204}}.

\bibitem{BouYanTA}
R.~Bousso and I.-S. Yang {\em (to appear)}  .

\end{thebibliography}\endgroup
\end{document}